\documentclass{article}
\usepackage{waspaa21,amsmath,graphicx,url,times}
\usepackage{color}
\usepackage{etoolbox,siunitx}
\robustify\bfseries
\usepackage{amssymb}
\usepackage{booktabs}
\def\H{{\mathsf H}}
\def\T{{\mathsf T}}
\def\CC{{\mathbb C}}

\usepackage{cite}
\usepackage{makecell}
\let\OLDthebibliography\thebibliography
\renewcommand\thebibliography[1]{
  \OLDthebibliography{#1}
  \setlength{\parskip}{0.5pt}
  \setlength{\itemsep}{1pt plus 0.3ex}
}

\title{Convolutive Prediction for Reverberant Speech Separation}

\name{Zhong-Qiu Wang, Gordon Wichern, Jonathan Le Roux}
\address{Mitsubishi Electric Research Laboratories (MERL), USA \\
{\small\texttt{wang.zhongqiu41@gmail.com,\{wichern,leroux\}@merl.com}}
}

\begin{document}

\ninept
\maketitle

\setlength{\abovedisplayskip}{4pt}
\setlength{\belowdisplayskip}{4pt}

\begin{sloppy}

\begin{abstract}

We investigate the effectiveness of convolutive prediction, a novel formulation of linear prediction for speech dereverberation, for speaker separation in reverberant conditions.
The key idea is to first use a deep neural network (DNN) to estimate the direct-path signal of each speaker, and then identify delayed and decayed copies of the estimated direct-path signal.
Such copies are likely due to reverberation, and can be directly removed for dereverberation or used as extra features for another DNN to perform better dereverberation and separation.
To identify such copies, we solve a linear regression problem per frequency efficiently in the time-frequency (T-F) domain to estimate the underlying room impulse response (RIR).
In the multi-channel extension, we perform minimum variance distortionless response (MVDR) beamforming on the outputs of convolutive prediction.
The beamforming and dereverberation results are used as extra features for a second DNN to perform better separation and dereverberation.
State-of-the-art results are obtained on the SMS-WSJ corpus.

\end{abstract}


\vspace{-0.05cm}
\section{Introduction}
\label{sec:intro}
\vspace{-0.05cm}

Dramatic progress has been made on speaker separation in anechoic conditions, since the inventions of deep clustering and permutation invariant training (PIT) \cite{R.Hershey2016, Isik2016, Kolbak2017}.
Room reverberation is pervasive in real-world applications, and speaker separation in reverberant conditions remains a challenging task.
In reverberant rooms, speech signals propagate in the air and are reflected many times inside the room.
The signal captured by far-field microphones contains an infinite number of delayed and decayed copies of the dry source signals.
Reverberation degrades speech intelligibility and quality, and is harmful to modern automatic speech recognition (ASR) systems.
Simultaneous speaker separation and dereverberation is a challenging task, as it is difficult to differentiate and separate the direct-path signal from its copies, especially when reverberation is strong and when there are multiple speakers.

Weighted prediction error (WPE) \cite{Nakatani2010} is so far the most popular dereverberation algorithm.
It estimates the late reverberation at the current frame by applying a linear filter to past observations, and then subtracts the estimate from the mixture for dereverberation.
The filter is estimated alternately with the target power spectral density (PSD).
WPE is found to introduce little speech distortion, leading to consistent improvements in many robust ASR studies \cite{Kinoshita2016, Boeddecker2018}.
Other conventional approaches for dereverberation include computing a Wiener filter based on estimated reverberation time \cite{Habets2009} or by using the estimated PSD of late reverberation \cite{Braun2018}.

Another popular approach for dereverberation is based on supervised deep learning, where DNNs are trained to estimate the direct-path signal from the mixture in a data-driven way \cite{WDL2018}.
The rationale is that clean speech exhibits strong spectral-temporal patterns, which can be modelled by powerful learning machines such as DNNs.
DNNs were initially used in the magnitude domain to predict T-F masks or target magnitudes \cite{Han2015}.
In the DNN-WPE algorithm \cite{Kinoshita2017}, the target PSD in WPE is estimated by DNNs so that the linear filter can have a closed-form solution and the iterative procedure is avoided.
Riding on the advance of deep learning, many subsequent DNN-based studies \cite{Luo2020, Wang2020b, Zhao2020, J.Borgstrom2020} have focused on designing advanced DNN architectures to predict target speech based on end-to-end training in the complex T-F or time domain.
However, there are few studies explicitly exploiting the linear-filter structure of reverberation, i.e., the fact that reverberation results from a linear convolution between an RIR and a dry source signal.
Intuitively, such a structure could be used as a regularizer for better dereverberation.

In this context, our study investigates the combination of linear prediction and deep learning to exploit the linear-filter structure for single- and multi-channel reverberant speaker separation and dereverberation, where we first use a DNN to estimate the direct-path signal of each speaker and then identify its delayed and decayed copies as the outcome of a forward filtering step.
Such copies are used to compute extra features to train another DNN for better dereverberation and separation.
We name the proposed dereverberation algorithm forward convolutive prediction (FCP), and compare its performance on reverberant speaker separation with DNN-WPE \cite{Kinoshita2017}, which implicitly exploits the linear-filter structure through inverse filtering.

\vspace{-0.05cm}
\section{System Overview}
\label{sec:overview}
\vspace{-0.05cm}

Given a $C$-speaker mixture recorded in a noisy-reverberant environment by a $P$-microphone array, the physical model in the short-time Fourier transform (STFT) domain can be formulated as
\begin{align} 
	\mathbf{Y}(t,f) &= \sum\nolimits_{c=1}^{C} \mathbf{X}(c,t,f)+\mathbf{V}(t,f) \nonumber \\
	&= \sum\nolimits_{c=1}^{C} \big(\mathbf{S}(c,t,f)+\mathbf{H}(c,t,f)\big)+\mathbf{V}(t,f), \label{eq:phymodel_freq}
\end{align}
where $\mathbf{Y}(t,f)$, $\mathbf{V}(t,f)$, $\mathbf{X}(c,t,f)$, $\mathbf{S}(c,t,f)$ and $\mathbf{H}(c,t,f)\in \CC^{P}$ respectively denote the STFT vectors of the mixture, noise, reverberant speech, direct and non-direct signals of speaker $c$, at time $t$ and frequency $f$.
The noise in this study is assumed to be a weak stationary noise.
Our study aims at recovering each speaker's direct-path signal captured at a reference microphone $q$, i.e., $S_q(c)$, based on $\mathbf{Y}$.
Variables without $t$ and $f$ refer to the corresponding spectrogram. To avoid clutter, we drop $f$ from the equations whenever computations are performed independently per frequency.

Figure \ref{systemfigure} illustrates the proposed two-DNN system.
The first DNN is trained using utterance-wise permutation invariant training (uPIT) \cite{Isik2016,Kolbak2017} to estimate the direct-path signal of each speaker at each microphone, denoted as $\hat{S}_q^{\text{DNN}_1}(c)$.
The target estimates are used to compute statistics for dereverberation based on convolutive prediction, and MVDR beamforming.
The second DNN takes in the outputs of the first DNN as well as the beamforming and dereverberation steps as features to enhance each target speaker.
Both DNNs are trained using single- or multi-microphone complex spectral mapping \cite{Wang2020c,Wang2020e}, where we predict the real and imaginary (RI) components of target speech based on the RI components of the stacked input signals.
DNN$_1$ is trained using the ``PIT+sumPIT'' loss proposed in \cite{Wang2021FCPjournal}, and DNN$_2$ is trained using either the ``RI'' loss or the ``RI+Mag'' loss presented in \cite{Wang2021FCPjournal}.
This two-DNN system is built upon a recent state-of-the-art speaker separation system, MISO-BF-MISO \cite{Wang2020c}, where an MVDR module is used in between the two networks.
The major contributions of this study are the introduction of a novel dereverberation module in between the two DNNs, and its integration with beamforming.

\begin{figure}[t]
  \centering
  \centerline{\includegraphics[width=\columnwidth]{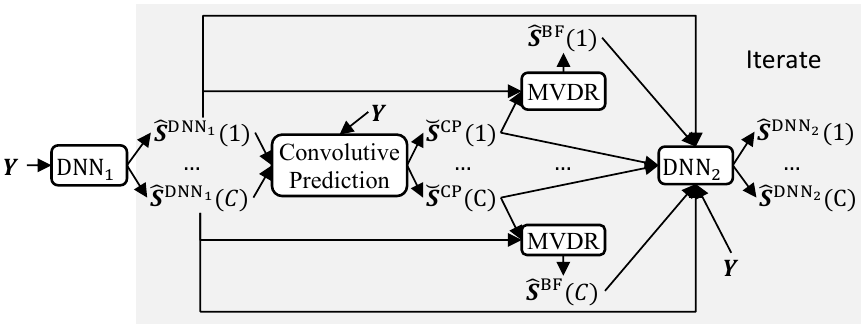}}\vspace{-0.45cm}
  \caption{System illustration.}\vspace{-0.6cm}
  \label{systemfigure}
\end{figure}

\vspace{-0.25cm}
\section{WPE and DNN-WPE}
\label{sec:wpe}
\vspace{-0.2cm}

This section reviews WPE \cite{Nakatani2010} and DNN-WPE \cite{Kinoshita2017}, and points out their strength and weakness.
While WPE was originally designed for single-speaker dereverberation, we adapt it to perform dereverberation in the context of reverberant speaker separation by estimating a dereverberation filter for each speaker, rather than estimating a single filter to dereverberate the mixture.
Since each speaker is convolved with a different RIR, it seems reasonable to estimate a dereverberation filter for each speaker.

WPE \cite{Nakatani2010} computes a $K$-tap inverse linear filter per frequency to estimate the late reverberation at the current frame from the past observations.
The estimated late reverberation
is then subtracted from the mixture for dereverberation, i.e.,
\begin{align}\label{wperesult}
\breve{S}_q^{\text{WPE}}(c,t)=Y_q(t)-\hat{\mathbf{g}}_q(c)^{\H}\widetilde{\mathbf{Y}}(t-\Delta),
\end{align}
where $\hat{\mathbf{g}}_q(c)\in \CC^{KP}$ is a $KP$-dimensional filter, $\Delta$ ($\geq 1$) a prediction delay, and
$\widetilde{\mathbf{Y}}(t)=[\mathbf{Y}(t)^\T,\dots,\mathbf{Y}(t-K+1)^\T]^\T$.
Assuming that the estimated target speech follows a zero-mean Gaussian distribution with time-varying PSD $\lambda_q(c,t)$, i.e., $\breve{S}_q^{\text{WPE}}(c,t) \sim \mathcal{N}\big(0, \lambda_q(c,t)\big)$, and based on maximum likelihood estimation, WPE computes the filter through the minimization problem
\begin{align}%
\underset{\substack{\mathbf{g}_q(c), \lambda_q(c)}}{{\text{argmin}}} \sum\nolimits_{t} \frac{|Y_q(t)-\mathbf{g}_q(c)^{\H}\ \widetilde{\mathbf{Y}}(t-\Delta)|^2}{\lambda_q(c,t)}+\text{log}\lambda_q(c,t),
\end{align}
where $|\cdot|$ computes magnitude.
This objective does not have a closed-form solution. An iterative algorithm is proposed in \cite{Nakatani2010} to alternately estimate $\mathbf{g}_q(c)$ and $\lambda(c,t)$.

Given a typical 32 ms STFT window size and an 8 ms hop size, $\Delta$ is usually set by default, or tuned through a validation set, to 3 or 4, because smaller $\Delta$ makes $Y_q(t)$ and $\widetilde{\mathbf{Y}}(t-\Delta)$ share time-domain signals due to the overlap between nearby frames, and will more likely result in target cancellation.
However, a large $\Delta$ would likely limit WPE’s capability to reduce early reflections.
Our work aims at removing both early reflections and late reverberation.

In the subsequent DNN-WPE algorithm \cite{Kinoshita2017}, $\lambda$ is estimated by a magnitude-domain DNN model, and 
the simplified objective is
\begin{align}\label{dnnwpe}
\underset{\mathbf{g}_q(c)}{{\text{argmin}}} \sum\nolimits_{t} \frac{|Y_q(t)-\mathbf{g}_q(c)^{\H}\ \widetilde{\mathbf{Y}}(t-\Delta)|^2}{\hat{\lambda}_q(c,t)}.
\end{align}
$\hat{\lambda}_q(c,t,f)=\text{max}(\varepsilon \text{max}(|\hat{S}_q^{\text{DNN}_b}(c)|^2),|\hat{S}_q^{\text{DNN}_b}(c,t,f)|^2)$, where $b\in \{1,2\}$ indicates one of the two DNNs, $\text{max}(\cdot)$ extracts the maximum value of a spectrogram, $\text{max}(\cdot,\cdot)$ returns the larger of two values, and $\varepsilon$ is a floor value to avoid putting too much weight on silent T-F units.
This quadratic objective has a closed-form solution.
The dereverberation result $\breve{S}_q^{\text{DNN-WPE}}$ is computed using Eq.~(\ref{wperesult}).

Compared with WPE, DNN-WPE leverages the modeling power of a DNN on magnitude-domain speech patterns to improve PSD estimation.
It makes WPE suitable for online dereverberation \cite{Heymann2018b} and makes the joint training of WPE with other DNN modules practical \cite{Heymann2019, ZhangWangyou2020, Zhang2021}.
Motivated by DNN-WPE, we explore other ways of using DNN-provided statistics for linear prediction.

One insight for potential improvement is that DNN-WPE only utilizes DNN-estimated target magnitude (i.e., by using it to compute $\hat{\lambda}$).
Many recent studies have suggested that phase estimation can also be improved by using deep learning \cite{Williamson2016, Wang2018d, Luo2019, Wang2020b, Wang2019}.
Our proposed algorithm leverages both magnitude and phase estimated by a DNN for filter estimation.
Another insight for potential improvement is that the linear filtering in WPE is applied to the mixture, which consists of multiple sources. The computed filter could be biased towards dereverberating higher-energy sources.

\vspace{-0.25cm}
\section{Proposed Algorithms}
\label{sec:proposed}
\vspace{-0.2cm}

To tackle these problems, we propose DNN-supported FCP for dereverberation in the context of reverberant speaker separation, and analyze its robustness to interferences.
We then present a multi-step FCP extension, and combine FCP with MVDR for multi-channel processing.
A post-filtering technique is presented at last.

\vspace{-0.2cm}
\subsection{Forward Convolutive Prediction (FCP)}
\vspace{-0.1cm}

In FCP, we approximate the mixture $Y_q(t)$ by forward filtering of the target speech $\hat{S}_q^{\text{DNN}_b}$ estimated by the DNN.
The filter is obtained by solving the minimization problem
\begin{align}\label{complexproj2}
\underset{\mathbf{g}_q(c)}{{\text{argmin}}} \sum\nolimits_{t} \frac{|Y_q(t)-\mathbf{g}_q(c)^{\H}\ \widetilde{\hat{\mathbf{S}}}{}_q^{\text{DNN}_b}(c,t)|^2}{\hat{\eta}_q(c,t)},
\end{align}
where $\widetilde{\hat{\mathbf{S}}}{}_q^{\text{DNN}_b}(c,t)\!=\![\hat{S}_q^{\text{DNN}_b}(c,t),\dots,\hat{S}_q^{\text{DNN}_b}(c,t-K+1)]^\T$ and $\hat{\eta}_q(c,t,f)=\text{max}(\varepsilon \text{max}(|Y_q|^2),|Y_q(t,f)|^2)$.
While DNN-WPE stems from a probabilistic model, we here introduce the denominator merely as a weighting that can balance the contribution of T-F units with diverse energy levels.
The objective to minimize is quadratic and a closed-form solution exists.
Note that among all the signals in the mixture, $\hat{S}_q^{\text{DNN}_b}(c)$, if sufficiently accurate, is expected to only correlate with the reverberant speech of speaker $c$.
Therefore, $\hat{\mathbf{g}}_q(c)^{\H}\widetilde{\hat{\mathbf{S}}}{}_q^{\text{DNN}_b}(c,t)$ can only approximate $X_q(c,t)$ for a time-invariant $\hat{\mathbf{g}}_q(c)$.
The dereverberation result is obtained as
\begin{align}
\breve{S}_q^{\text{FCP}}(c,t) = Y_q(t)-\big(\hat{\mathbf{g}}_q(c)^{\H}\widetilde{\hat{\mathbf{S}}}{}_q^{\text{DNN}_b}(c,t)-\hat{S}_q^{\text{DNN}_b}(c,t)\big),
\label{eq:s_fcp}
\end{align}
where the subtracted term from $Y_q(t)$ is considered as the estimated reverberation of speaker $c$.
Note that $\breve{S}_q^{\text{FCP}}(c)$ still contains the reverberant signals of the other sources, as Eq. (\ref{eq:s_fcp}) only reduces the reverberation of a target speaker from the mixture and preserves everything else.
We can reduce the reverberation of all the target speakers by combining their FCP results (denoted as cFCP):
\begin{align}
\!\breve{S}_q^{\text{cFCP}}\!(c,t) \!=\! Y_q(t)\!-\!\sum_{c'}\!\big(\hat{\mathbf{g}}_q(c')^{\H}\widetilde{\hat{\mathbf{S}}}{}_q^{\text{DNN}_b}(c',t)\!-\!\hat{S}_q^{\text{DNN}_b}\!(c',t)\big).
\label{cFCP}
\end{align}

Compared with (\ref{dnnwpe}), Eq.~(\ref{complexproj2}) may better reduce early reflections because a prediction delay is not necessary.
In addition, it can utilize both magnitude and phase estimated by DNNs for linear prediction.

\vspace{-0.2cm}
\subsection{Robustness of WPE and FCP to Interference}
\vspace{-0.1cm}

Eq.~(\ref{complexproj2}) may lead to better filter estimation than (\ref{dnnwpe}) for the target speaker when interferences are present.
To see this, we equivalently reformulate Eq.~(\ref{complexproj2}) in terms of $X_q$: denoting $\mathbf{N}(c)=\mathbf{Y}-\mathbf{X}(c)$,
\begin{align}\label{complexproj2robust}
&\underset{\mathbf{g}_q(c)}{{\text{argmin}}} \sum\nolimits_{t} \frac{|X_q(c,t)+N_q(c,t)-\mathbf{g}_q(c)^{\H}\ \widetilde{\hat{\mathbf{S}}}{}_q^{\text{DNN}_b}(c,t)|^2}{\hat{\eta}_q(c,t)} \nonumber \\
&= %
\underset{\mathbf{g}_q(c)}{{\text{argmin}}} \sum\nolimits_{t} \frac{|X_q(c,t)-\mathbf{g}_q(c)^{\H}\ \widetilde{\hat{\mathbf{S}}}{}_q^{\text{DNN}_b}(t)|^2+|N_q(c,t)|^2}{\hat{\eta}_q(c,t)} \nonumber \\
&= %
\underset{\mathbf{g}_q(c)}{{\text{argmin}}} \sum\nolimits_{t} \frac{|X_q(c,t)-\mathbf{g}_q(c)^{\H}\ \widetilde{\hat{\mathbf{S}}}{}_q^{\text{DNN}_b}(c,t)|^2}{\hat{\eta}_q(c,t)},
\end{align}
where the analysis assumes that $\hat{S}_q^{\text{DNN}_b}$ and $X_q(c)$ are uncorrelated with $N_q(c)$, meaning that
\begin{align}
\sum\nolimits_{t} \frac{N_q(c,t)^{\H}\Big(X_q(c,t)-\mathbf{g}_q(c)^{\H}\ \widetilde{\hat{\mathbf{S}}}{}_q^{\text{DNN}_b}(c,t)\Big)}{\hat{\eta}_q(c,t)}\approx 0.
\end{align}
This derivation suggests that FCP essentially estimates the filter using $\hat{S}_q^{\text{DNN}_b}$ and $X_q(c)$, between which a linear-filter structure exists.
This could produce a good filter estimate for each target speaker, even if the mixture includes competing speakers and noise.

A similar derivation for Eq.~(\ref{dnnwpe}) leads to
\begin{align}\label{wperobust}
&\underset{\mathbf{g}_q(c)}{{\text{argmin}}} \!\sum_{t} \!\frac{{
\footnotesize
|X_q(c,t)\!+\!\!N_q(c,t) \!-\! \mathbf{g}_q(c)^{\H}\! \big(\widetilde{\mathbf{X}}(c,t\!-\!\Delta)\!+\!\widetilde{\mathbf{N}}(c,t\!-\!\Delta)\big)|^2}}
{\hat{\lambda}_q(c,t)} \nonumber \\
&= %
\underset{\mathbf{g}_q(c)}{{\text{argmin}}}
\big(
\sum\nolimits_{t} \frac{|X_q(c,t)-\mathbf{g}_q(c)^{\H}\widetilde{\mathbf{X}}(c,t-\Delta)|^2}{\hat{\lambda}_q(c,t)} \nonumber \\
&\hspace{1.3cm}
+ \sum\nolimits_{t} \frac{|N_q(c,t)-\mathbf{g}_q(c)^{\H}\widetilde{\mathbf{N}}(c,t-\Delta)|^2}
{\hat{\lambda}_q(c,t)}
\big),
\end{align}
where $\widetilde{\mathbf{X}}(c,t)$ and $\widetilde{\mathbf{N}}(c,t)$ are defined similarly to $\widetilde{\mathbf{Y}}(t)$.
This derivation suggests that WPE aims at dereverberating the target speaker and non-target sources using a single filter.
This could be problematic when non-target sources are present and the number of sources exceeds the number of microphones (i.e., in under-determined cases), because the filter would also need to reduce the reverberation of non-target sources rather than focusing on dereverberating the target speaker.
When they are strong, in under-determined cases the loss on non-target sources could dominate the numerator, and the resulting filter may be biased towards dereverberating higher-energy sources.
In contrast, Eq.~(\ref{complexproj2}) of FCP aims at only removing the reverberation related to a target speaker.
This is particularly useful in multi-speaker separation, because each target speaker is convolved with a different RIR and it is thus reasonable to compute a different dereverberation filter for each speaker.
This also means that our current method does not aim at using linear prediction to reduce the reverberation of non-target sources such as multi-source environmental noises, as it would require estimating each anechoic noise source, which is very difficult \cite{Kavalerov2019}.
We think this is fine because we have a second DNN to leverage convolutive-prediction outputs for further enhancement.

\vspace{-0.2cm}
\subsection{Multi-Step FCP (msFCP)}
\vspace{-0.1cm}

For Eq.~(\ref{complexproj2}) to boil down to (\ref{complexproj2robust}), $\hat{S}_q^{\text{DNN}_b}$ needs to be sufficiently accurate; otherwise, linearly filtering it to approximate $Y_q$ would not be able to approximate $X_q(c)$.
Ideally, we would want to estimate the filter using Eq.~(\ref{complexproj2robust}), but $X_q(c)$ has to be estimated beforehand.
One way is to train a separate DNN or add an output in our first DNN to estimate it, at the cost of increased DNN complexity.
Considering that $V$ is a weak stationary noise in this study, we propose multi-step FCP, where we remove from $Y_q$ the reverberation estimated in the previous step to refine the target used in FCP.
More specifically, in step one we apply Eq.~(\ref{complexproj2}) to estimate an FCP filter $\hat{\mathbf{g}}_q(c;1)$ for each speaker $c$.
At step $i>1$, we compute the filter $\hat{\mathbf{g}}_q(c;i)$ as 
\begin{align}\label{complexproj2multistep}
\underset{\mathbf{g}_q(c;i)}{{\text{argmin}}} \sum\nolimits_{t} \frac{|\hat{Z}_q(c,t;i\text{-}1)-\mathbf{g}_q(c;i)^{\H}\ \widetilde{\hat{\mathbf{S}}}{}_q^{\text{DNN}_b}(c,t)|^2}{\hat{\tau}_q(c,t;i\text{-}1)},
\end{align}
where $\hat{Z}_q(c,t;i\text{-}1)=Y_q(t)-\sum_{c'\neq c}\hat{\mathbf{g}}_q(c';i\text{-}1)^{\H}\ \widetilde{\hat{\mathbf{S}}}{}_q^{\text{DNN}_b}(c',t)$ can be considered as an estimation of $X_q(c,t)$, and $\hat{\tau}_q(c,t,f;i\text{-}1)=\text{max}(\varepsilon \text{max}(|\hat{Z}_q(c;i\text{-}1)|^2),|\hat{Z}_q(c,t,f;i\text{-}1)|^2)$.
The dereverberation result is obtained as
\begin{align}
\breve{S}_q^{\text{msFCP}}\!(c,t;i) \!=\! \hat{Z}_q(c,t;i\text{-}1)\!-\! \big(\hat{\mathbf{g}}_q(c;i)^{\H}\widetilde{\hat{\mathbf{S}}}{}_q^{\text{DNN}_b}\!(c,t)\!-\!\hat{S}_q^{\text{DNN}_b}\!(c,t)\big).
\label{msFCP}
\end{align}
Two steps are applied in our experiments. 
Note that different from $\breve{S}_q^{\text{FCP}}(c)$ and $\breve{S}_q^{\text{cFCP}}(c)$, $\breve{S}_q^{\text{msFCP}}(c)$ is expected to only contain the anechoic speech of speaker $c$ and reverberant noises.

\vspace{-0.2cm}
\subsection{Combining FCP with MVDR Beamforming}
\vspace{-0.1cm}

Following \cite{Drude2018, Nakatani2020, Boeddeker2020}, we then apply MVDR beamforming to dereverberation outputs to further improve separation and dereverberation.
The target and non-target covariance matrices, $\hat{\mathbf{\Phi}}(c)$ and $\hat{\mathbf{\Phi}}(\lnot c)$, are computed as
\begin{align}\label{covariancematrix}
\hat{\mathbf{\Phi}}(c) &= \sum\nolimits_t \hat{\mathbf{S}}^{\text{DNN}_b}(c,t)\hat{\mathbf{S}}^{\text{DNN}_b}(c,t)^{\H}, \\
\hat{\mathbf{\Phi}}(\lnot c) &= \sum\nolimits_t \hat{\mathbf{U}}^{\text{DNN}_b}(\lnot c,t)\hat{\mathbf{U}}^{\text{DNN}_b}(\lnot c,t)^{\H}, \\
    \hat{\mathbf{U}}^{\text{DNN}_b}(\lnot c) &= \breve{\mathbf{S}}^{\text{Dereverb}}(c)-\hat{\mathbf{S}}^{\text{DNN}_b}(c), \label{nontarget1}
\end{align}
where $\breve{\mathbf{S}}^{\text{Dereverb}}(c)$ denotes the results of FCP, cFCP, msFCP, or DNN-WPE.
Following \cite{Yoshioka2015, Zhang2017a}, the steering vector $\hat{\mathbf{d}}(c)$ of speaker $c$ is computed as the principal eigenvector of $\hat{\mathbf{\Phi}}(c)$.
Designating microphone $q$ as the reference, an MVDR beamformer is computed as
$\hat{\mathbf{w}}(c;q) = \frac{\hat{\mathbf{\Phi}}(\lnot c)^{-1} \hat{\mathbf{d}}(c)}{\hat{\mathbf{d}}(c)^{\H} \hat{\mathbf{\Phi}}(\lnot c)^{-1} \hat{\mathbf{d}}(c)} \hat{d}_q^{*}(c)$,
where $(\cdot)^{*}$ computes the complex conjugate, and beamforming results are computed as
\begin{align}\label{bf}
\hat{S}_q^{\text{BF}}(c,t) = \hat{\mathbf{w}}(c;q)^{\H}\breve{\mathbf{S}}^{\text{Dereverb}}(c,t).
\end{align}
Alternatively, we can compute $\hat{\mathbf{U}}^{\text{DNN}_b}(\lnot c)$ using
\begin{align}\label{nontarget2}
\hat{\mathbf{U}}^{\text{DNN}_b}(\lnot c)=\mathbf{Y}-\hat{\mathbf{S}}^{\text{DNN}_b}(c),
\end{align}
and apply the resulting beamformer to the mixture.

\vspace{-0.2cm}
\subsection{Post-Filtering}
\vspace{-0.1cm}

FCP exploits the linear-filter structure in reverberation, and MVDR leverages the linear spatial information among multiple microphones.
Both of them could provide information complementary to plain DNN-based end-to-end dereverberation and separation.
We hence combine their outputs with the mixture as input features to train DNN$_2$ to enhance each target speaker.

As $\hat{\mathbf{S}}^{\text{DNN}_2}$ is likely better than $\hat{\mathbf{S}}^{\text{DNN}_1}$, at run time we use it to do another pass of FCP and MVDR, and feed the new FCP and MVDR results to DNN$_2$ to estimate each speaker again.

\vspace{-0.25cm}
\section{Experiments}
\label{sec:experiments}
\vspace{-0.2cm}

\subsection{Dataset and System Configurations}
\vspace{-0.1cm}

We validate the proposed algorithms using the six-channel SMS-WSJ dataset \cite{Drude2019}, which contains 33,561, 982, and 1,332 simulated reverberant two-speaker mixtures for training, validation, and testing, respectively.
The speaker-to-array distance is sampled from the  range $[1.0, 2.0]$~m, and the T60 is drawn from the range $[0.2, 0.5]$~s.
A weak white noise is added to simulate microphone noise. The energy level between the sum of the reverberant target speech signals and the noise is sampled from the range $[20, 30]$~dB.
The sampling rate is 8~kHz.
We use the direct sound, obtained by setting T60 to $0$ s, as the labels for model training and perform joint dereverberation, separation, and denoising.
We consider monaural separation, where the first microphone is used for model training and testing, and two-channel separation using the first and fourth microphones.
We use the default ASR backend provided with SMS-WSJ for recognition, trained on single-speaker reverberant speech. %

For STFT, the window size is 32 ms and hop size 8 ms.
After cross-validation, $K$ is set to 37 and $\Delta$ to 3 for DNN-WPE, $K$ is set to 40 for FCP, and $\varepsilon$ is tuned to $0.001$ for $\hat{\lambda}$, $\hat{\eta}$, and $\hat{\tau}$.
The DNN architectures follow \cite{Wang2020c}.
Scale-invariant signal-to-distortion ratio (SI-SDR) \cite{LeRoux2019}, perceptual evaluation of speech quality (PESQ) \cite{P862.1} and word error rate (WER) are used as the evaluation metrics.

\vspace{-0.2cm}
\subsection{Results}
\vspace{-0.1cm}

Table \ref{results1ch} reports monaural (1ch) results.
We only go over the SI-SDR numbers, as similar trends are observed for PESQ and WER.
For now, we only look at the entries where $\text{DNN}_2$ is trained using the ``RI'' loss in \cite{Wang2021FCPjournal}.
$\text{DNN}_1$, a uPIT network, improves the performance from $-5.5$ to $6.1$ dB.
$\text{DNN}_1$+DNN$_2$, which combines the mixture with the outputs of $\text{DNN}_1$ to train an enhancement network (DNN$_2$) to enhance each speaker, improves the performance to $9.8$ dB.
We can also include the outcomes of WPE or FCP, computed based on $\text{DNN}_1$ outputs, to train DNN$_2$.
Among them, $\text{DNN}_1$+msFCP+DNN$_2$ shows 
the best performance at $12.2$ dB.
Doing another pass of msFCP and running DNN$_2$ one more time, denoted as $\text{DNN}_1$+(msFCP+DNN$_2$)$\times$2, improves the performance from $12.2$ to $14.0$ dB.
In contrast, doing another pass on WPE only improves the performance slightly.
This is likely because DNN$_2$ can produce better magnitude and phase than $\text{DNN}_1$, and FCP can leverage these better magnitude and phase for better reverberation estimation, while WPE only leverages the magnitude.
The $14.0$ dB result is substantially better than a recent complex spectral mapping based system (SISO) \cite{Wang2020c} and DPRNN-TasNet \cite{Luo2020}, both of which are popular end-to-end approaches in speaker separation.

Table \ref{results2ch} presents two-microphone results.
Using two-channel uPIT, $\text{DNN}_1$ obtains $8.5$ dB.
Plain DNN stacking, $\text{DNN}_1$+DNN$_2$, gets to $12.2$ dB.
Including MVDR results computed using $\text{DNN}_1$ outputs to train DNN$_2$, denoted as $\text{DNN}_1$+MVDR+DNN$_2$, improves the performance from $12.2$ to $12.8$ dB.
This MVDR is computed by using Eq.~(\ref{nontarget2}) and the beamformer is applied to the mixture.
We can include the outcomes of FCP or WPE computed based on $\text{DNN}_1$ outputs to train DNN$_2$.
Among them, $\text{DNN}_1$+MVDR+msFCP+DNN$_2$ performs slightly better.
We can also apply MVDR beamforming to the results of FCP or WPE %
rather than to the mixture (denoted as, for example, $\text{DNN}_1$+msFCP\_MVDR+msFCP+DNN$_2$).
This leads to better performance.
By doing one more pass of msFCP\_MVDR and msFCP, we get our best score, $16.1$ dB.
This result is substantially better than two popular end-to-end systems, FasNet-TAC \cite{Luo2020d} and multi-channel ConvTasNet \cite{ZhangJisi2020}, and a recent MISO-BF-MISO system \cite{Wang2020c}, which is essentially the same as $\text{DNN}_1$+MVDR+DNN$_2$.

Training DNN$_2$ with the ``RI+Mag'' loss presented in \cite{Wang2021FCPjournal} produces better PESQ and WER, and slightly worse SI-SDR for the DNN$_1$+msFCP+DNN$_2$ system in Table~\ref{results1ch} and the DNN$_1$+msFCP\_MVDR+msFCP+DNN$_2$ system in Table~\ref{results2ch}.
This observation aligns with the findings in \cite{Wang2021compensation}.

\begin{table}[t]
\scriptsize
\centering
  \sisetup{table-format=2.2,round-mode=places,round-precision=2,table-number-alignment = center,detect-weight=true,detect-inline-weight=math}
\caption{\footnotesize{SI-SDR (dB), PESQ and WER (\%) results on SMS-WSJ (1ch).}}
\label{results1ch}
\begin{tabular}{lcS[table-format=2.1,round-precision=1]S[table-format=1.2,round-precision=2]S[table-format=2.1,round-precision=1]}
\toprule
{\bf Approaches} & {\bf DNN$_2$ Loss} & {\bf SI-SDR} & {\bf PESQ} & {\bf WER} \\
\midrule
Unprocessed & - & -5.5 & 1.50 & 78.42 \\ %
DNN$_1$ & - & 6.1 & 2.17 & 38.42 \\ %
DNN$_1$+DNN$_2$ & RI & 9.8 & 2.64 & 23.39 \\
\midrule
DNN$_1$+WPE+DNN$_2$ & RI & 11.0 & 2.81 & 18.82 \\ %
DNN$_1$+FCP+DNN$_2$ & RI & 12.0 & 2.89 & 18.26 \\ %
DNN$_1$+cFCP+DNN$_2$ & RI & 11.3 & 2.78 & 20.47\\ %
DNN$_1$+msFCP+DNN$_2$ & RI & 12.2 & 3.04 & 16.04 \\
DNN$_1$+msFCP+DNN$_2$ & RI+Mag & 11.6 & 3.25 & 13.22 \\
\midrule
DNN$_{1}$+(WPE+DNN$_2$)$\times2$ & RI & 11.4 & 2.88 & 18.23 \\ %
DNN$_1$+(FCP+DNN$_2$)$\times2$ & RI & 13.0 & 3.00 & 16.33 \\ %
DNN$_1$+(cFCP+DNN$_2$)$\times2$ & RI & 12.4 & 2.84 & 20.68 \\ %
DNN$_1$+(msFCP+DNN$_2$)$\times2$ & RI & \bfseries 14.0 & 3.30 & 13.84 \\
DNN$_1$+(msFCP+DNN$_2$)$\times2$ & RI+Mag & 13.4 & \bfseries 3.41 & \bfseries 10.93 \\
\midrule
SISO \cite{Wang2020c} & - & 5.1 & 2.40 & 28.28 \\ %
DPRNN-TasNet \cite{Luo2020} & - & 6.5 & 2.28 & 38.12 \\
\bottomrule
\end{tabular}\vspace{-0.5cm}
\end{table}

\begin{table}[t]
\scriptsize
\centering
  \sisetup{table-format=2.1,round-mode=places,round-precision=2,table-number-alignment = center,detect-weight=true,detect-inline-weight=math}
\caption{\footnotesize{SI-SDR (dB), PESQ and WER (\%) results on SMS-WSJ (2ch).}}
\label{results2ch}
\setlength{\tabcolsep}{2pt}
\begin{tabular}{lcS[table-format=2.1,round-precision=1]S[table-format=1.2,round-precision=2]S[table-format=2.1,round-precision=1]}
\toprule
{\bf Approaches} & {\bf DNN$_2$ Loss} & {\bf SI-SDR} & {\bf PESQ} & {\bf WER} \\
\midrule
Unprocessed & - & -5.5 & 1.50 & 78.42 \\ %
DNN$_1$ & - & 8.5 & 2.53 & 27.12 \\ %
DNN$_1$+DNN$_2$ & RI & 12.2 & 3.00 & 15.04 \\ %
DNN$_1$+MVDR+DNN$_2$ & RI & 12.8 & 3.16 & 13.78 \\
\midrule
DNN$_1$+MVDR+WPE+DNN$_2$ & RI & 13.6 & 3.25 & 12.60 \\ %
DNN$_1$+MVDR+FCP+DNN$_2$ & RI & 13.9 & 3.26 & 13.16 \\ %
DNN$_1$+MVDR+cFCP+DNN$_2$ & RI & 14.1 & 3.35 & 11.74 \\ %
DNN$_1$+MVDR+msFCP+DNN$_2$ & RI & 14.1 & 3.37 & 11.73 \\
\midrule
DNN$_1$+WPE\_MVDR+WPE+DNN$_2$ & RI & 14.3 & 3.37 & 11.57 \\ %
DNN$_1$+FCP\_MVDR+FCP+DNN$_2$ & RI & 14.3 & 3.35 & 12.17 \\ %
DNN$_1$+cFCP\_MVDR+cFCP+DNN$_2$ & RI & 14.4 & 3.38 & 11.80 \\ %
DNN$_1$+msFCP\_MVDR+msFCP+DNN$_2$ & RI & 14.5 & 3.44 & 11.15 \\
DNN$_1$+msFCP\_MVDR+msFCP+DNN$_2$ & RI+Mag & 14.2 & 3.63 & 9.12 \\
\midrule
DNN$_1$+(WPE\_MVDR+WPE+DNN$_2$)$\times2$ & RI & 14.4 & 3.37 & 11.72 \\ %
DNN$_1$+(FCP\_MVDR+FCP+DNN$_2$)$\times2$ & RI & 15.4 & 3.46 & 11.71 \\ %
DNN$_1$+(cFCP\_MVDR+cFCP+DNN$_2$)$\times2$ & RI & 15.5 & 3.49 & 11.17 \\ %
DNN$_1$+(msFCP\_MVDR+msFCP+DNN$_2$)$\times2$ & RI & \bfseries 16.1 & 3.65 & 10.04 \\
DNN$_1$+(msFCP\_MVDR+msFCP+DNN$_2$)$\times2$ & RI+Mag & 15.8 & \bfseries 3.71 & \bfseries 8.55 \\
\midrule
FasNet-TAC \cite{Luo2020d} & - & 6.9 & 2.31 & 34.85 \\ %
Multi-channel ConvTasNet \cite{ZhangJisi2020} & - & 5.8 & 2.60 & 45.72\\ %
MISO-BF-MISO \cite{Wang2020c} & - & 12.3 & 3.39 & 11.39 \\ 
\bottomrule
\end{tabular}\vspace{-0.45cm}
\end{table}

\vspace{-0.25cm}
\section{Conclusion}
\label{sec:conclusions}
\vspace{-0.2cm}

We have proposed convolutive prediction for reverberant speech separation and dereverberation, and combined it with beamforming in the multi-channel case.
Evaluation results show that the proposed convolutive prediction leads to better separation and ASR performance than DNN-WPE in the context of a state-of-the-art two-DNN speech separation system, in both single- and multi-channel scenarios. %
In closing, we emphasize that the linear-filter structure in reverberation provides an informative cue for dereverberation, and explicitly exploiting it could be an important step towards solving the cocktail party problem in realistic conditions.

\bibliographystyle{IEEEtran}
{\footnotesize\bibliography{refs21}}

\begin{thebibliography}{10}
\providecommand{\url}[1]{#1}
\def\UrlFont{\rmfamily}
\providecommand{\newblock}{\relax}
\providecommand{\bibinfo}[2]{#2}
\providecommand\BIBentrySTDinterwordspacing{\spaceskip=0pt\relax}
\providecommand\BIBentryALTinterwordstretchfactor{4}
\providecommand\BIBentryALTinterwordspacing{\spaceskip=\fontdimen2\font plus
\BIBentryALTinterwordstretchfactor\fontdimen3\font minus
  \fontdimen4\font\relax}
\providecommand\BIBforeignlanguage[2]{{%
\expandafter\ifx\csname l@#1\endcsname\relax
\typeout{** WARNING: IEEEtran.bst: No hyphenation pattern has been}%
\typeout{** loaded for the language `#1'. Using the pattern for}%
\typeout{** the default language instead.}%
\else
\language=\csname l@#1\endcsname
\fi
#2}}

\bibitem{R.Hershey2016}
J.~R. Hershey, Z.~Chen, J.~{Le Roux}, and S.~Watanabe, ``{Deep Clustering:
  Discriminative Embeddings for Segmentation and Separation},'' in \emph{Proc.
  ICASSP}, 2016, pp. 31--35.

\bibitem{Isik2016}
Y.~Isik, J.~{Le Roux}, Z.~Chen, S.~Watanabe, and J.~R. Hershey,
  ``{Single-Channel Multi-Speaker Separation using Deep Clustering},'' in
  \emph{Proc. Interspeech}, 2016, pp. 545--549.

\bibitem{Kolbak2017}
M.~Kolb{\ae}k, D.~Yu, Z.-H. Tan, and J.~Jensen, ``{Multitalker Speech
  Separation with Utterance-Level Permutation Invariant Training of Deep
  Recurrent Neural Networks},'' \emph{IEEE/ACM Trans. Audio, Speech, Lang.
  Process.}, vol.~25, no.~10, pp. 1901--1913, 2017.

\bibitem{Nakatani2010}
T.~Nakatani, T.~Yoshioka, K.~Kinoshita, M.~Miyoshi, and B.-H. Juang, ``{Speech
  Dereverberation Based on Variance-Normalized Delayed Linear Prediction},''
  \emph{IEEE Trans. Audio, Speech, Lang. Process.}, vol.~18, no.~7, pp.
  1717--1731, 2010.

\bibitem{Kinoshita2016}
K.~Kinoshita, M.~Delcroix, S.~Gannot, E.~A. Emanu{\"{e}}l, R.~Haeb-Umbach,
  W.~Kellermann, V.~Leutnant, R.~Maas, T.~Nakatani, B.~Raj, A.~Sehr, and
  T.~Yoshioka, ``{A Summary of The REVERB Challenge: State-of-The-Art and
  Remaining Challenges in Reverberant Speech Processing Research},''
  \emph{EURASIP J. Adv. Signal Process.}, vol. 2016, no.~1, pp. 1--19, 2016.

\bibitem{Boeddecker2018}
C.~Boeddecker, J.~Heitkaemper, J.~Schmalenstroeer, L.~Drude, J.~Heymann, and
  R.~Haeb-Umbach, ``{Front-End Processing for The CHiME-5 Dinner Party
  Scenario},'' in \emph{Proc. CHiME-5}, 2018, pp. 35--40.

\bibitem{Habets2009}
E.~{A. P. Habets}, S.~Gannot, and I.~Cohen, ``{Late Reverberant Spectral
  Variance Estimation Based on A Statistical Model},'' \emph{IEEE Signal
  Process. Lett.}, vol.~16, no.~9, pp. 770--773, 2009.

\bibitem{Braun2018}
S.~Braun, A.~Kuklasinski, O.~Schwartz, O.~Thiergart, E.~{A. P. Habets},
  S.~Gannot, S.~Doclo, and J.~Jensen, ``{Evaluation and Comparison of Late
  Reverberation Power Spectral Density Estimators},'' \emph{IEEE/ACM Trans.
  Audio, Speech, Lang. Process.}, vol.~26, no.~6, pp. 1052--1067, 2018.

\bibitem{WDL2018}
D.~Wang and J.~Chen, ``{Supervised Speech Separation Based on Deep Learning: An
  Overview},'' \emph{IEEE/ACM Trans. Audio, Speech, Lang. Process.}, vol.~26,
  no.~10, pp. 1702--1726, 2018.

\bibitem{Han2015}
K.~Han, Y.~Wang, D.~Wang, W.~{S. Woods}, I.~Merks, and T.~Zhang, ``{Learning
  Spectral Mapping for Speech Dereverberation and Denoising},'' \emph{IEEE/ACM
  Trans. Audio, Speech, Lang. Process.}, vol.~23, no.~6, pp. 982--992, 2015.

\bibitem{Kinoshita2017}
K.~Kinoshita, M.~Delcroix, H.~Kwon, T.~Mori, and T.~Nakatani, ``{Neural
  Network-Based Spectrum Estimation for Online WPE Dereverberation},'' in
  \emph{Proc. Interspeech}, 2017, pp. 384--388.

\bibitem{Luo2020}
Y.~Luo, Z.~Chen, and T.~Yoshioka, ``{Dual-Path RNN: Efficient Long Sequence
  Modeling for Time-Domain Single-Channel Speech Separation},'' in \emph{Proc.
  ICASSP}, 2020, pp. 46--50.

\bibitem{Wang2020b}
Z.-Q. Wang and D.~Wang, ``{Deep Learning Based Target Cancellation for Speech
  Dereverberation},'' \emph{IEEE/ACM Trans. Audio, Speech, Lang. Process.},
  vol.~28, pp. 941--950, 2020.

\bibitem{Zhao2020}
Y.~Zhao, D.~Wang, B.~Xu, and T.~Zhang, ``{Monaural Speech Dereverberation using
  Temporal Convolutional Networks with Self Attention},'' \emph{IEEE/ACM Trans.
  Audio, Speech, Lang. Process.}, vol.~28, pp. 1598--1607, 2020.

\bibitem{J.Borgstrom2020}
B.~{J. Borgstrom} and M.~{S. Brandstein}, ``{The Speech Enhancement via
  Attention Masking Network (SEAMNET): An End-to-end System for Joint
  Suppression of Noise and Reverberation},'' \emph{IEEE/ACM Trans. Audio,
  Speech, Lang. Process.}, 2020.

\bibitem{Wang2020c}
Z.-Q. Wang, P.~Wang, and D.~Wang, ``{Multi-Microphone Complex Spectral Mapping
  for Utterance-Wise and Continuous Speaker Separation},'' \emph{IEEE/ACM
  Trans. Audio, Speech, Lang. Process.}, vol.~29, pp. 2001--2014, 2021.

\bibitem{Wang2020e}
Z.-Q. Wang and D.~Wang, ``{Multi-Microphone Complex Spectral Mapping for Speech
  Dereverberation},'' in \emph{Proc. ICASSP}, 2020, pp. 486--490.

\bibitem{Wang2021FCPjournal}
Z.-Q. Wang, G.~Wichern, and J.~{Le Roux}, ``{Convolutive Prediction for
  Monaural Speech Dereverberation and Noisy-Reverberant Speaker Separation},''
  \emph{in submission}, 2021.

\bibitem{Heymann2018b}
J.~Heymann, L.~Drude, R.~Haeb-Umbach, K.~Kinoshita, and T.~Nakatani,
  ``{Frame-Online DNN-WPE Dereverberation},'' in \emph{Proc. IWAENC}, 2018, pp.
  466--470.

\bibitem{Heymann2019}
------, ``{Joint Optimization of Neural Network-Based WPE Dereverberation and
  Acoustic Model for Robust Online ASR},'' in \emph{Proc. ICASSP}, 2019, pp.
  6655--6659.

\bibitem{ZhangWangyou2020}
W.~Zhang, A.~S. Subramanian, X.~Chang, S.~Watanabe, and Y.~Qian, ``{End-to-End
  Far-Field Speech Recognition with Unified Dereverberation and Beamforming},''
  in \emph{Proc. Interspeech}, 2020, pp. 324--328.

\bibitem{Zhang2021}
W.~Zhang, C.~Boeddeker, S.~Watanabe, T.~Nakatani, M.~Delcroix, K.~Kinoshita,
  T.~Ochiai, N.~Kamo, R.~Haeb-Umbach, and Y.~Qian, ``{End-to-End
  Dereverberation, Beamforming, and Speech Recognition with Improved Numerical
  Stability and Advanced Frontend},'' in \emph{Proc. ICASSP}, 2021.

\bibitem{Williamson2016}
D.~S. Williamson, Y.~Wang, and D.~Wang, ``{Complex Ratio Masking for Monaural
  Speech Separation},'' \emph{IEEE/ACM Trans. Audio, Speech, Lang. Process.},
  pp. 483--492, 2016.

\bibitem{Wang2018d}
Z.-Q. Wang, J.~{Le Roux}, D.~Wang, and J.~R. Hershey, ``{End-to-End Speech
  Separation with Unfolded Iterative Phase Reconstruction},'' in \emph{Proc.
  Interspeech}, 2018, pp. 2708--2712.

\bibitem{Luo2019}
Y.~Luo and N.~Mesgarani, ``{Conv-TasNet: Surpassing Ideal Time-Frequency
  Magnitude Masking for Speech Separation},'' \emph{IEEE/ACM Trans. Audio,
  Speech, Lang. Process.}, vol.~27, no.~8, pp. 1256--1266, 2019.

\bibitem{Wang2019}
Z.-Q. Wang, K.~Tan, and D.~Wang, ``{Deep Learning Based Phase Reconstruction
  for Speaker Separation: A Trigonometric Perspective},'' in \emph{Proc.
  ICASSP}, 2019, pp. 71--75.

\bibitem{Kavalerov2019}
I.~Kavalerov, S.~Wisdom, H.~Erdogan, B.~Patton, K.~Wilson, J.~{Le Roux}, and
  J.~R. {Hershey}, ``{Universal Sound Separation},'' in \emph{Proc. WASPAA},
  2019, pp. 175--179.

\bibitem{Drude2018}
L.~Drude, C.~Boeddeker, J.~Heymann, R.~Haeb-Umbach, K.~Kinoshita, M.~Delcroix,
  and T.~Nakatani, ``{Integrating Neural Network Based Beamforming and Weighted
  Prediction Error Dereverberation},'' in \emph{Proc. Interspeech}, 2018, pp.
  3043--3047.

\bibitem{Nakatani2020}
T.~Nakatani, C.~Boeddeker, K.~Kinoshita, R.~Ikeshita, M.~Delcroix, and
  R.~Haeb-Umbach, ``{Jointly Optimal Denoising, Dereverberation, and Source
  Separation},'' \emph{IEEE/ACM Trans. Audio, Speech, Lang. Process.}, vol.~28,
  pp. 2267--2282, 2020.

\bibitem{Boeddeker2020}
C.~Boeddeker, T.~Nakatani, K.~Kinoshita, and R.~Haeb-Umbach, ``{Jointly Optimal
  Dereverberation and Beamforming},'' in \emph{Proc. ICASSP}, 2020, pp.
  216--220.

\bibitem{Yoshioka2015}
T.~Yoshioka, N.~Ito, M.~Delcroix, A.~Ogawa, K.~Kinoshita, M.~Fujimoto, C.~Yu,
  W.~J. Fabian, M.~Espi, T.~Higuchi, S.~Araki, and T.~Nakatani, ``{The NTT
  CHiME-3 System: Advances in Speech Enhancement and Recognition for Mobile
  Multi-Microphone Devices},'' in \emph{Proc. ASRU}, 2015, pp. 436--443.

\bibitem{Zhang2017a}
X.~Zhang, Z.-Q. Wang, and D.~Wang, ``{A Speech Enhancement Algorithm by
  Iterating Single- and Multi-Microphone Processing and Its Application to
  Robust ASR},'' in \emph{Proc. ICASSP}, 2017, pp. 276--280.

\bibitem{Drude2019}
L.~Drude, J.~Heitkaemper, C.~Boeddeker, and R.~Haeb-Umbach, ``{SMS-WSJ}:
  Database, performance measures, and baseline recipe for multi-channel source
  separation and recognition,'' \emph{arXiv preprint arXiv:1910.13934}, 2019.

\bibitem{LeRoux2019}
J.~{Le Roux}, S.~Wisdom, H.~Erdogan, and J.~R. {Hershey}, ``{SDR - Half-Baked
  or Well Done?}'' in \emph{Proc. ICASSP}, 2019, pp. 626--630.

\bibitem{P862.1}
\BIBentryALTinterwordspacing
``{P.862.1 : Mapping function for transforming P.862 raw result scores to
  MOS-LQO},'' 2003. [Online]. Available:
  \url{https://www.itu.int/rec/T-REC-P.862.1-200311-I/en}
\BIBentrySTDinterwordspacing

\bibitem{Luo2020d}
Y.~Luo, Z.~Chen, N.~Mesgarani, and T.~Yoshioka, ``{End-to-End Microphone
  Permutation and Number Invariant Multi-Channel Speech Separation},'' in
  \emph{Proc. ICASSP}, 2020, pp. 6394--6398.

\bibitem{ZhangJisi2020}
J.~Zhang, C.~Zorila, R.~Doddipatla, and J.~Barker, ``{On End-to-End
  Multi-Channel Time Domain Speech Separation in Reverberant Environments},''
  in \emph{Proc. ICASSP}, 2020, pp. 6389--6393.

\bibitem{Wang2021compensation}
Z.-Q. Wang, G.~Wichern, and J.~{Le Roux}, ``{On The Compensation Between
  Magnitude and Phase in Speech Separation},'' \emph{in submission}, 2021.

\end{thebibliography}

\end{sloppy}
\end{document}